# Bandwidth Slicing to Boost Federated Learning in Edge Computing

Jun Li[1], Xiaoman Shen[2], Lei Chen[3] and Jiajia Chen[1]

[1]*Department of Electronic Engineering, Chalmers University of Technology, Göteborg, Sweden (Email: jiajiac@chalmers.se)*
[2]*Zhejiang University, Hangzhou, China;* [3]*Research Institutes of Sweden, Göteborg, Sweden*

**Abstract:** Bandwidth slicing is introduced to support federated learning in edge computing to assure low communication delay for training traffic. Results reveal that bandwidth slicing significantly improves training efficiency while achieving good learning accuracy.

## 1. Introduction

Edge computing (EC) has been regarded as a promising technology to enable time-critical services (e.g., augmented reality, industry automation), in which computing and storage functions can be delivered to end users at the edge of networks. Meanwhile, many services deployed at the EC nodes can be greatly enhanced by using machine learning technologies (e.g., deep learning and reinforcement learning). For example, a deep learning based object recognition technique can deliver accurate results for face and gesture recognition [1]. On the other hand, these deep learning based applications highly depend on the quality and quantity of dada sets for model training. Due to privacy preserving, it is not always practical to aggregate/share all the data from various learners from distinct locations or different organizations at a centralized data center.

Federated learning (FL) is a new paradigm of distributed learning, where clients, like end users and/or EC nodes can collaboratively learn a shared model while keeping all the data locally. During the training process of the FL, each client needs to periodically transmit its local model parameters to the centralized parameter server (CPS), where a set of global model parameters are updated according to aggregation strategies such as federated averaging algorithm (FedAvg) [2], and then the CPS sends the global model parameters to each client for its local model updates. It often needs many rounds to achieve optimal learning performance. The synchronous training, which requests the model updates to be carried out within a fixed time period, typically runs more efficiently than the asynchronous one [2]. On the other hand, the synchronous training puts high requirements on communication networks, to make sure model updates can be done quickly to achieve high training efficiency. However, the slow clients, also referred to as stragglers, affect synchronization time. The amount of training traffic generated by the model updates per round can be huge. For example, the training traffic for a convolutional neural network (CNN) is up to tens of Mbits. A deadline-driven client selection scheme [4] filters the staggers and hence reduces the training time. However, the stragglers' contribution to the training process is ignored, and thereby the learning accuracy may be degraded significantly.

In this regard, this paper introduces bandwidth slicing to boost federated learning by reducing synchronization time during the training. We consider the edge computing scenario, where passive optical network (PON) as a promising technology used to support the communications among EC nodes that aggregate traffic from end users [5]. EC nodes associated with optical network units (ONUs) transmit the training data to the CPS co-located with the optical line terminal (OLT) at central office (CO). The other traffic for mobile backhaul and broadband access can co-exist in the same PON. Our results show that the proposed bandwidth slicing significantly outperforms the benchmark that simply follows first come first served (FCFS) queuing policy, saving training time more than 30%, even under a high traffic load (e.g., 0.8).

## 2. Bandwidth slicing for the federated learning in edge computing

The EC nodes can be equipped with standalone severs (e.g., cloudlet, roadside unit), which are directly connected with ONUs with high-speed Ethernet interface. The EC nodes can be deployed at, for instance, residential houses for smart home, shopping malls for self-service shopping and roadside units for intelligent transport systems, which may need run various learning tasks. At the CO, the high-layer functions of the OLT can be supported by a general computing platform to facilitate model update/aggregation. One slice can be assigned to one learning task, therefore, multiple learning tasks can be supported by allocating multiple slices.

Figure 1 shows the proposed network slicing mechanism for the FL in edge computing over a PON, where we consider one OLT/CPS and three ONUs/EC nodes as an example. The upper part of the time line for the OLT/CPS represents the computation procedure, while the lower part represents the communication procedure. For the time line of each ONU/EC node, the upper part represents the communication procedure, while the lower part is for the computation procedure. For each client, involved in the FL task, the synchronization time per round consists of global model downloading, local model training, local model uploading and model aggregation [3]. At the beginning of each round, the OLT broadcasts the global model updated in the previous round to the ONUs/EC nodes via a certain amount of reserved downlink bandwidth, and the communication time for $ONU_i/EC_i$ for the global model downloading is denoted as $T_i^{DL}$. Then, each involved EC node runs training by employing the downloaded global model and its own local data. The time used for completing the local model training at $ONU_i/EC_i$ is referred to as $T_i^{UD}$. The local model training depends on many factors, such as complexity of models, resources (e.g., CPU, memory), data sizes, and model hyper parameters (e.g., batch size, local epochs). For one FL task, even with the same model and the same hyper parameters, the heterogeneity of EC nodes cannot be ignored. The edge computing architecture is often heterogeneous, where various EC nodes usually have different amounts of the local computing resources and data, and therefore $T_i^{UD}$ ($i \in V$, where $V$ is the set of ONUs/EC nodes) may vary significantly. After completing the local training, the ONUs that are associated with the involved EC nodes send the model updates ($M_i^{UD}$) to the CPS in their allocated time slots which are determined by the proposed BS algorithm (see Algorithm). In this way, the communication time of local model uploading ($T_i^{UL}$) can be assured. After gathering updates from all clients, the CPS aggregates local models and updates the global model. Then, one synchronization round finishes. $T_a$ denotes the time for the model aggregation at the CPS, which depends on the complexity of the chosen aggregation algorithm and available computing power at the CPS. AvgFed [2] is employed in this paper, which is a very simple averaging algorithm but efficient. We assume the CPS has powerful computing resource and the running time of AvgFed (i.e., $T_a$) can be ignored.

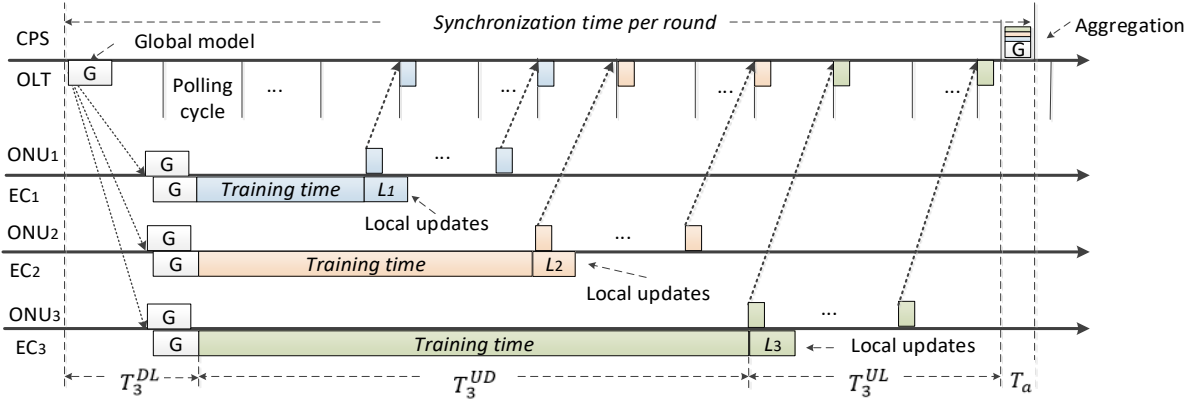

Fig.1. Illustration of network slicing in edge computing for federated learning

The proprosed BS algorithm is triggered only when new clients join or leave the FL task. Once the BS algorithm is triggered, the OLT updates the corresponding slice for the FL task by using the information of all involved cleints ($\Phi$). Note each EC node can involve many cleints for the same FL task. If there are new clients that join in the FL procedure, the first synchronization round that the new clidents join needs to get $\Phi$. As shown in Fig. 1, $T_i^{UD}$ ($i \in V$) are highly heterogenerous. Correspondly, $\Delta$ denotes the sum of $T_i^{DL}$ and $T_i^{UD}$ ($i \in V$). By utilizing the heteriority of $\Delta$, the training traffic can be transmitted within the time gap ($\tau$) between its maximum value plus $\nabla$ ($T^{max} + \nabla$) and its minimum value ($T^{min}$), where $\nabla$ denotes the time to transmit the training traffic with the latest arrival time and can be estimated based on the distance between the ONUs and the OLT. Capacity ($B$), starting time ($t_s$) and ending time ($t_e$) of the slice can be caculated, correspondingly. Once one slice is created, due to local training $h$ ($H > h \geq 1$) syhcronization rounds may be experienced before the slice is assigned to the training data, where $H$ is the total number of the syncronization rounds for the training task. As mentioned earlier, the shorter the synchronization round is, the shorter the training time can be achieved. In the proposed BS algorithm, the threshold is denoated by $T^{round}$, which is a fixed value set by the CPS. To receive the local models from all the clients involved in the training processing, it is obvious that $T^{round}$ should be not less than the sum of $T_i^{DL}$, $T_i^{UD}$, $T_i^{UL}$ and $T_a$, where $i \in V$. Besides, $B$ should not be larger than the uplink capaciy ($C$). Otherwise, $T_i^{UD}$ has to be reduced to realize the required synchronization time. Once $B$ is known for the current slice, the OLT schedules a fixed time slot for each ONU. As bandwidth in PONs are allocated to different ONUs periodically (i.e., in each polling cycle), thus the allocated slice can be further mapped into each polling cycle, as shown in Fig. 1.

**Algorithm: Bandwidth slicing algorithm (BS)**
**Input:** $\Phi\{T_i^{UD}; M_i^{UD}; i \in V\}$; $t_{current}$; $T^{round}$; $C$
**Output**: $S\{t_s; t_e; B\}$
1. **for** $i = 1:1:length(V)$
2. $\quad \Delta \leftarrow T_i^{UD} + T_i^{DL}$
3. **end for**
4. **Sort** ($\Delta$) according to the descend of $T_i^{UD}$
5. $T^{max} \leftarrow Max(\Delta) + \nabla$
6. $T^{min} \leftarrow Min(\Delta)$
7. $\tau \leftarrow T^{max} - T^{min}$
8. $B \leftarrow Max(\sum_{i \in V} M_i^{UD}/\tau, C)$
9. $t_e \leftarrow t_{current} + T^{max} + h \times T^{round}$;
10. $t_s \leftarrow t_{current} + T^{min} + h \times T^{round}$;
11. $h \in \{1,2,3,...H-1\}$

## 3. Performance evaluation

A home-made simulator written by Python is developed to evaluate the performance of the proposed BS algorithm for the FL in edge computing. The dataset of FEMNIST [6] is used for model training, where a CNN model with two 5x5 convolution layers is employed and the FedAvg algorithm [2] has been chosen for parameter aggregation at the CPS. The traffic generated by the CNN model update is 26.416 Mbits for each client. The hyper parameters (e.g., learning rate, batch size) refer to [6]. The background traffic follows Poisson distribution, which together with training traffic determines the total traffic load. The simulations are performed on a computer with Intel Core i7-6700CPU@3.4GHx8, Memory 31.3GiB, and Linux 18.04.3 LTS. A time division multiplexing PON with 128 ONUs is used, in which the line rate for both upstream and downstream is set at 10 Gbps. The distance between the OLT and ONUs is set to 20 km. Each ONU can involve up to 24 clients and connects to only one EC node, where one FL task is implemented.

In Fig. 2(a), the learning accuracy becomes statured with a large number of rounds. The more the clients are involved in the FL task, the more the rounds are needed to get the statured accuracy. For the specific case studied in this paper, the highest accuracy is about 68% when the percentage of the involved clients is 10%, while it increases to 82% with 100% involvement of the clients. Meanwhile, giving a certain accuracy level, the training efficiency is determined by the synchronization time per round. For each round of the FL, the synchronization time includes both communication time and computation time. As shown in Fig. 2(b), the blue curve represents the percentage of involving clients as a function of the computation time (i.e., $T_i^{UD}$), which stands for the minimum synchronization time per round (i.e., without considering any communication delay) ranging from 1s to 5s. We consider the benchmark that simply follows FCFS queuing policy for bandwidth allocation in both upstream and downstream, where the high traffic load leads to a large synchronization time. When the traffic load is 0.3, the synchronization time per round including both computation and communication increases to 6.7s for 100% involvement of the clients when employing the FCFS. When the total traffic load increases to 0.8, the maximum synchronization time continues to increase (~8s). In comparison, the proposed BS algorithm reserves the dedicated bandwidth to the FL task, thus the corresponding communication delay reduces significantly and is not affected by the total traffic load. To achieve the highest accuracy, 82% in this case, where 100% involvement of the clients is needed, the proposed BS algorithm can save 36% of the training time compared to the FCFS when the total traffic load is 0.8.

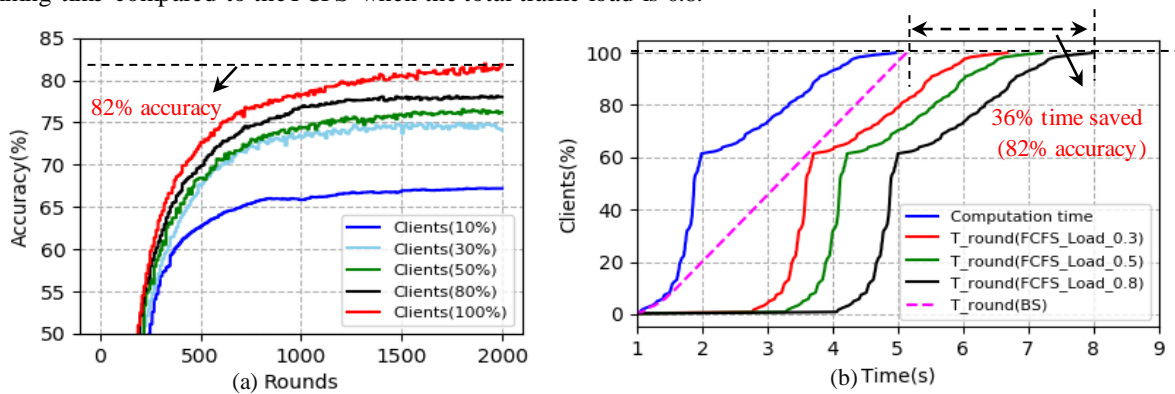

Figure 2. (a) Learning accuracy vs. number of rounds, and (b) percentage of the involved clients vs. the synchronization time.

## 4. Conclusions

This paper introduces bandwidth slicing for federated learning in edge computing with heterogenrity of clients' computation time. Bandwidth slicing is able to reserve network resource dedicated to the learning task, which can well address the inssues brought by stragglers during the training process. Simulation results show that the FL training effiicency can be significantly improved while achieving the same level of learning accuracy. For the specific FL task studied in this paper, the traning time can be saved up to 36 % to achieve the maximum learning accuracy.

**Acknowledgement:** This work is supported in part by SJTU State Key Laboratory of Advanced Optical Communication System and Networks Open Project 2018GZKF03001, Swedish Research Council (VR) project 2016-04489 "Go-iData", Swedish Foundation for Strategic Research (SSF), and Chalmers ICT-seed grant.